\documentclass[prb,twocolumn,showpacs]{revtex4}
\usepackage[dvips]{graphicx}
\usepackage{epsfig}
\usepackage{color}
\usepackage[normalem]{ulem}
\usepackage{amsmath,amsfonts,amssymb,bm}%,ulem}
\begin{document}
\title{Low temperature 1/f noise in microwave dielectric constant of amorphous dielectrics in Josephson qubits}
\author{Alexander L. Burin}
\affiliation{Department of Chemistry, Tulane University, New
Orleans, LA 70118, USA}
\author{Moshe Schechter, Shlomi Matityahu}
\affiliation{Department of Physics, Ben-Gurion University of the
Negev, Beer Sheva 84105, Israel}

% \author{B. I. Shklovskii}
% \affiliation{William P. Fine Institute of Theoretical Physics,
% School of Physics and Astronomy, University of Minnesota,
% Minneapolis, MN 55455, USA}
%\author{Arkady A. Kurnosov}
%\affiliation{A. F. Ioffe  Physico-Technical Institute of Russian
%Academy of Sciences, 194021 St. Petersburg, Russia}
%\author{V. Vinokur}
%\affiliation{Argonne National Laboratory, 9700 S. Cass Av.,
%Argonne, IL 60439, USA}
\date{\today}
\begin{abstract}
The accurate analytical solution for the low temperature $1/f$
noise in a microwave dielectric constant of amorphous films
containing tunneling two-level systems (TLSs) is derived within
the standard tunneling model including the weak dipolar or elastic
TLS-TLS interactions. The results are consistent with the recent
experimental investigations of $1/f$ noise in Josephson junction
qubits including the power law increase of the noise amplitude
with decreasing temperature at low temperatures $T<0.1$K. The long
time correlations needed for $1/f$ noise are provided by the
logarithmic broadening of TLS absorption resonances with time due
to their interaction with neighboring TLSs. The noise behavior at
higher temperatures $T>0.1$K and its possible sensitivity to
quasi-particle excitations are discussed.
\end{abstract}

\maketitle
%\small{
%\yg{We still have space. I believe that the following remains
%  to be done:
%\begin{itemize}
%\item Agree on numerical estimates.
%\end{itemize}
%}

%\section{Introduction}

%\label{intr}

$1/f$ noise exists in a variety of physical
systems,\cite{classik1,DattaRev88,WeissmanRev,ShMKogan} and it
dramatically restricts performance of modern electronic and
quantum nanodevices.\cite{McCammon,McCammon1,Martinis} The inverse
frequency dependence of the noise power spectral density,
$S^{}_{xx}(f)\propto 1/f$, is a consequence of a logarithmically
slow relaxation. The slow dynamics is often associated with the
random ensemble of fluctuators possessing a logarithmically
uniform spectrum of relaxation
times.\cite{WeissmanRev,ShMKogan,abNoise,Aleiner,Clare0} Such
fluctuators do exist in amorphous solids in the form of universal
tunneling two-level systems (TLSs).\cite{AHVP}
%Particularly these TLSs can be responsible for the noise in superconducting quantum bits.\cite{Martinis}
With the advent of superconducting quantum bits (qubits)
based on Josephson junctions,\cite{ReviewGen,Martinis} a comprehensive
study of the noise properties due to TLSs has become crucial for
the achievement of high-fidelity quantum computation.
TLSs are ubiquitous, appearing in wiring dielectrics, Josephson
junction barriers and other disordered insulating regions.
The deleterious effects of the coupling of TLSs to
the qubit are associated with the resonant absorption of the
qubit energy by TLSs in the microwave frequency range,
as well as by $1/f$ noise in the microwave dielectric
constant,\cite{Martinis} resulting in qubit decoherence.

Dissipation and $1/f$ noise in
superconducting resonators due to TLSs have been extensively
studied.\cite{Noise1,Noise1_5,Noise2,NoiseRecent,NoiseKevin}
Recently, it was found that the $1/f$ noise amplitude
in high quality superconducting resonators increases with
decreasing temperature as $T^{-1-\eta}$, with $\eta \approx
0.3$.\cite{NoiseRecent} This dependence was considered as being
incompatible with the standard tunneling model
(STM).\cite{NoiseRecent,Faoro3}

A qualitative theoretical model interpreting this recent
experiment has been proposed in Refs.\ \onlinecite{NoiseRecent}
and \onlinecite{Faoro3}, suggesting an
energy-dependent TLSs density of states (DOS), $g(E)\approx
E^{\eta}$, in contrast with the standard TLS model suggesting a
constant DOS.\cite{AHVP} The authors used their assumption to
interpret the additional exponent $\eta$ in the noise temperature
dependence and the anomalous temperature dependence of TLSs
decoherence
%\sout{and relaxation}
rate discovered earlier using a single TLS
spectroscopy.\cite{UstinovRecent} This assumption conflicts with
earlier experimental data showing energy independence of the DOS,
leading to logarithmic temperature dependence of the dielectric
constant and sound velocity\cite{TLSDOS,Hunklinger} (see however
Ref.\ \onlinecite{UstinovNew}).

%Our solution explains the low temperature noise behavior in the regime where the thermal energy is much smaller than the field quantization energy $k_{B}T\ll\hbar\omega$ with.
In this paper we derive an exact analytical solution
for the low temperature $1/f$ noise in a microwave dielectric
constant of amorphous films, assuming a homogeneous DOS as in the
STM. This solution can be directly compared with the available
experimental data and used to quantitatively characterize TLSs
properties which is important for understanding of the nature of
TLSs and the reduction of their destructive effects 
in superconducting quantum devices. Particularly, we show that by
properly considering the logarithmic temperature dependence of the
noise power spectral density, the $T^{-1-\eta}$ dependence of the
amplitude of $1/f$ noise on temperature can be explained within
the assumptions of the STM and predict the noise dependence on the
external field amplitude. The derivation focuses on the relevant
regime of low temperatures, where the thermal energy is much
smaller than the field quantization energy, namely
$k^{}_{B}T\ll\hbar\omega$.

%On the other hand the qualitative analysis missed several
%important features compared to the analytical quantitative
%derivation suggested below.
%This includes the logarithmic temperature dependence of the noise
%power spectral density which can be used to interpret
%$T^{-1-\eta}$ dependence within the standard TLS model

%in the regime where the thermal energy is much smaller than the
%field quantization energy $k_{B}T\ll\hbar\omega$.
%The theory also predicts the stronger field dependencies of the
%noise power spectral density compared to Ref.\
%\onlinecite{Faoro3}.
%The proposed quantitative theory can be
%directly compared to the experimental data and used to extract the
%parameters of TLS so it is important for understanding the nature
%of TLS and reduction of their destructive effect in Josephson
%junction quantum devices. %Our solution explains the low temperature noise behavior   in the regime where the thermal energy is much smaller than the field quantization energy $k_{B}T\ll\hbar\omega$ with.

%Here we report the accurate analytical solution of the $1/f$ noise problem for microwave  dielectric constant using the standard TLS model \cite{AHVP} involving  weak TLS interactions,\cite{BlackHalperin} which results in the different noise behavior than the semi-qualitative solution.\cite{Faoro3} Our solution explains the low temperature noise behavior   in the regime where the thermal energy is much smaller than the field quantization energy $k_{B}T\ll\hbar\omega$.
At higher temperature, $k^{}_{B}T\geq\hbar\omega$, we argue that
TLSs based models cannot explain the noise
temperature dependence. An alternative noise source in this regime
is discussed by employing the possible contribution of
quasi-particles excitations in the superconductors. It is
demonstrated that such excitations may be responsible for the
anomalous temperature dependence of the relaxation and decoherence
rates of TLSs, observed in this regime.\cite{UstinovRecent}

% Expression for the time varying dielectric constant Eq. 3 with time dependent frequency, refer to adiabatic regime

Consider the TLSs contribution to the dielectric constant measured
using the input field, $F^{}_{AC}$, applied along the $z$ axis at
a microwave frequency $\omega$ and low temperature 
$k^{}_{B}T\ll\hbar\omega$. In this regime the resonant
contribution of TLSs to the real part of the dielectric constant
dominates and can be expressed in the
form\cite{Hunklinger,HunklingerNonLinAbs,Clare2,Faoro3}
\begin{eqnarray}
\frac{\epsilon^{}_{TLS}(t)}{\epsilon}=\frac{4\pi}{V\epsilon}\sum_{i}\tanh\left(\frac{E^{}_{i}}{2k^{}_{B}T}\right)
\nonumber\\
\times\frac{(E^{}_{i}(t)-\hbar\omega)p^{2}_{tr,iz}}{(E{}_{i}(t)-\hbar\omega)^2+T^{-2}_{2i}\left(1+p^{2}_{tr,iz}F^{2}_{AC}T^{}_{1i}T^{}_{2i}/\hbar^2\right)}.
\label{eq:deltaeps}
\end{eqnarray}
Here, $V$ and $\epsilon$ represent the sample volume and bare
dielectric constant. The summation is taken over all TLSs $i$
having transition dipole moments $\mathbf{p}^{}_{tr,i}$,
relaxation and decoherence times $T^{}_{1i}$ and $T^{}_{2i}$,
respectively, and time-dependent energies $E^{}_{i}(t)$. The $z$
component of the transition dipole moment is related to the
corresponding component of the TLS dipole moment, $p^{}_{iz}$, by
$p^{}_{tr,iz}=p^{}_{iz}\Delta^{}_{0i}/E^{}_{i}$,\cite{Hunklinger,Clare2}
where $\Delta^{}_{0i}$ is the tunneling amplitude of TLS $i$.
Following the STM,\cite{AHVP,Hunklinger} we assume that TLSs
possess the universal distribution
$P(\Delta^{}_{0},E)=P^{}_{0}E/(\Delta^{}_{0}\sqrt{E^2-\Delta^{2}_{0}})$
with respect to their energies and tunneling amplitudes, where
$P^{}_{0}$ is a material dependent constant.

The time dependence of the energies is induced by the spectral
diffusion caused by weak dipolar or elastic TLS-TLS
interaction.\cite{BlackHalperin}
%Assuming the contributing TLS to be close to resonance $|E_{i}(t)-\hbar\omega|\ll \hbar\omega$ one can ignore related time dependencies of other parameters.
The time dependence of the energy can be treated classically
because at the temperatures under consideration,
$k^{}_{B}T\ll\hbar\omega\approx E$, the equilibrium thermal
fluctuations can induce transitions of TLSs only with an
exponentially small probability $e^{-E/k^{}_{B}T}$.

Relaxation of TLSs is caused by phonon emission or absorption,
with the relaxation rate\cite{Hunklinger,BlackHalperin,abEcho}
\begin{eqnarray}
\frac{1}{T_{1i}}=A\left(\frac{\Delta^{}_{0i}}{E^{}_{i}}\right)^2
\frac{E^{3}_{i}}{k^{3}_{B}}\coth\left(\frac{E^{}_{i}}{2k^{}_{B}T}\right),
\label{eq:TLSRelRate}
\end{eqnarray}
where the proportionality constant $A \sim 10^8$s$^{-1}$K$^{-3}$
characterizes the TLS-phonon interaction.\cite{abEcho} The TLS
decoherence rate is composed of the contributions of relaxation
and pure phase decoherence,
$T^{-1}_{2i}=\left(2T^{}_{1i}\right)^{-1}+T^{-1}_{\varphi,i}$,
where the phase decoherence rate $T^{-1}_{\varphi,i}$ is
determined by the TLS spectral diffusion induced by its
interaction with neighboring thermal
TLSs, i.e.\ TLSs for which $E\approx\Delta^{}_{0}\approx k^{}_{B}T$.\cite{BlackHalperin} %One can define the phase decoherence rate using the two pulse echo decay $A_{2}(t)$ interpolated as $\exp\left(-\frac{t^2}{T_{2}^2}\right)$.
This rate is given by\cite{BlackHalperin,abEcho}
\begin{eqnarray}
\frac{1}{T^{}_{\varphi,i}}=\sqrt{40\frac{|\Delta^{}_{i}|}{E^{}_{i}}\frac{\chi
k^{}_{B}T\cdot AT^3}{\hbar}}, \label{eq:TLSPhDecRate}
\end{eqnarray}
where $\Delta^{}_{i}=\sqrt{E^{2}_{i}-\Delta^{2}_{0i}}$ is the TLS
asymmetry and the dimensionless constant
$\chi=P^{}_{0}U^{}_{0}\sim 5\cdot 10^{-4}$ represents the
universal product of the TLSs DOS and their $1/R^3$ interaction
strength.\cite{YuLeggett} The product $\chi k^{}_{B}T$ represents
the typical interaction
 with thermal TLSs and $AT^3$ represents the
relaxation rate of thermal TLSs. Although the use of
Eq.\ \eqref{eq:TLSPhDecRate} in the expression for the dielectric
constant [Eq.\ \eqref{eq:deltaeps}] has not been justified
theoretically, its approximate relevance was demonstrated
experimentally.\cite{HunklingerNonLinAbs} One should notice that
in the linear regime of weak external field the related term is
negligible in the noise spectral density (see Eq.\ \eqref{eq:noise4} below and estimates of parameters of this equation in the next two paragraphs). In the 
opposite regime of a very strong field the accuracy of our results is limited to the above assumption and may need a special consideration.
%not important for $1/f$ noise amplitude.

The noise is determined by the time-dependent correlation function
of the dielectric constant fluctuations
$S^{}_{xx}(t)=\langle\delta\epsilon^{}_{TLS}(t)\delta\epsilon_{TLS}(0)\rangle/\epsilon^2$,
where
$\delta\epsilon_{TLS}(t)\equiv\epsilon_{TLS}(t)-\langle\epsilon_{TLS}\rangle$.
Correlations between different resonant TLSs contributing to the
noise can be neglected because of the weakness of the TLS-TLS
interactions (see e.g., Ref.\ \onlinecite{YuLeggett}). The
correlation function can then be expressed as
\begin{widetext}
 \begin{eqnarray}
%\frac{\langle\delta\epsilon^{}_{TLS}(t)\delta\epsilon^{}_{TLS}(0)\rangle}{\epsilon^2}
S^{}_{xx}(t)= \frac{(4\pi)^2}{V^2
\epsilon^2}\sum_{i}p^{4}_{iz}\tanh^2\left(\frac{E^{}_{i}}{2k^{}_{B}T}\right)
\nonumber\\
\times\Bigg\langle\frac{E^{}_{i}(t)-\hbar\omega}{(E^{}_{i}(t)-\hbar\omega)^2+T^{-2}_{2i}\left(1+p^{2}_{iz}F^{2}_{AC}
T^{}_{1i}T^{}_{2i}/\hbar^2\right)}\cdot\frac{E_{i}(0)-\hbar\omega}{(E_{i}(0)-\hbar\omega)^2+T^{-2}_{2i}\left(1+p^{2}_{iz}F^{2}_{AC}T^{}_{1i}T^{}_{2i}/\hbar^2\right)}\Bigg\rangle.
\label{eq:Noise1}
\end{eqnarray}
\end{widetext}

The correlation function in Eq.\ \eqref{eq:Noise1} can be
evaluated by means of the distribution function $D(E|E_{0},t)$ for
the TLS energy $E$ at time $t$, assuming $E(0)=E^{}_{0}$ at
$t=0$.\cite{BlackHalperin} For $1/R^3$ interaction, this
distribution function is a Lorentzian
\begin{eqnarray}
D(E|E_{0},t)=\frac{1}{\pi}\frac{W(t)}{W^{2}(t)+(E-E^{}_{0})^2},
\label{eq:distr}
\end{eqnarray}
with a characteristic width $W(t)=W^{}_{0}(t)|\Delta|/E$, where
\begin{eqnarray}
W^{}_{0}(t)=\frac{\pi^2\chi
k^{}_{B}T}{3\hbar}\int_{0}^{\infty}\frac{dy}{\cosh^{2}y}\int_{0}^{1}\frac{1-e^{-rt/T^{}_{1}(y)}}{r}dr,
\label{eq:width}
\end{eqnarray}
and $T^{-1}_{1}(y)=8AT^3y^3\coth y$.

Low frequency $1/f$ noise is determined by long times $t\geq
1\text{s}\gg (AT^{3})^{-1}$,\onlinecite{NoiseRecent} where $(AT^{3})^{-1}$ estimates the minimum relaxation time of thermal TLS. In this limit the width of the distribution
$W^{}_{0}(t)$ grows logarithmically with time and can be
approximated as
\begin{eqnarray}
W^{}_{0}(t)\approx\frac{\pi^2}{3\hbar}\chi
k^{}_{B}T\ln\left(3.3\cdot AT^3 t\right).
\label{eq:width2}
\end{eqnarray}
This logarithmic increase of the resonance width is responsible
both for the appearance of $1/f$ noise and for the additional
exponent $\eta$ in the temperature dependence $T^{-1-\eta}$ of the
noise amplitude. This dependence has not been considered in Ref.\
\onlinecite{Faoro3}, leading the authors to the conclusion that
the experimental results are inconsistent with the STM.

%It  has been ignored in Ref. \cite{Faoro3} where the time has been replaced with some maximum relaxation time ($\Gamma_{min}^{-1}$, see Eq. (19), there). This can be justified if the noise frequency is smaller than the minimum decay rate which seem to conflict both with the experiment \cite{ab1} because logarithmic TLS relaxation has been observed for the longest measurements times extended to weeks and with the theory (see discussion below).
%As has been noticed in Ref. \cite{Faoro3} the noise is determined by the resonant TLSs $E\approx \hbar\omega$ and therefore
At low temperatures $W(t)\ll\hbar\omega$ and one can perform
averaging of the correlation function in Eq.\ \eqref{eq:Noise1}
over the distribution of energy fluctuations [Eq.\
\eqref{eq:distr}] and over initial TLS energies within the
resonant approximation. Straightforward evaluation of integrals
with respect to $E(t)$ and $E(0)$ yields
\begin{widetext}
\begin{eqnarray}
%\frac{\langle\delta\epsilon^{}_{TLS}(t)\delta\epsilon^{}_{TLS}(0)\rangle}{\epsilon^2}
S^{}_{xx}(t)=\tanh^2\left(\frac{\hbar\omega}{2k^{}_{B}T}\right)\frac{4\pi^3
P_{0}}{\hbar V\epsilon^2}
\int_{0}^{1}\frac{dx}{x(1-x^2)}\int_{0}^{1}dy\Bigg\langle\frac{p^{4}_{0}x^4y^4}
{3W^{}_{0}(t)+ \frac{T^{-1}_{2}(x)}{\sqrt{1-x^2}}
\sqrt{1+(xyp^{}_{0}F^{}_{AC})^2T^{}_{1}(x)T^{}_{2}(x)/\hbar^{2}}}\Bigg\rangle,
\label{eq:noise3}
\end{eqnarray}
\end{widetext}
where $x=\Delta^{}_{0}/\hbar\omega$ and $y=\cos\theta$. Averaging
in Eq.\ \eqref{eq:noise3} is made over absolute values of TLS
dipole moment, $p^{}_{0}$, forming an angle $\theta$ with the AC
field. In the numerical calculations below we assume $p^{}_{0}\sim
5$D, which is well justified experimentally.\cite{Martinis} The
relaxation rate $T^{-1}_{1}(x)$ is given by Eq.\
\eqref{eq:TLSRelRate} with $E^{}_{i}$ replaced by $\hbar\omega$
and $\Delta^{}_{0i}/E^{}_{i}$ replaced by $x$. Similarly,
$T^{-1}_{2}(x)=\left(2T^{}_{1}(x)\right)^{-1}+T^{-1}_{\varphi}(x)$,
where $T^{-1}_{\varphi}(x)$ is given by Eq.\
\eqref{eq:TLSPhDecRate} with $\Delta^{}_{i}/E^{}_{i}=\sqrt{1-(\Delta_{0i}/E_{i})^2}$ replaced by
$\sqrt{1-x^{2}}$.

The power spectral density of noise, $S^{}_{xx}(f)$, can be
evaluated as a Fourier transform of $S^{}_{xx}(t)$ [Eq.\
\eqref{eq:noise3}] in the low frequency limit $fT^{}_{1}$, $f/W\ll
1$. It has the pure $1/f$ spectrum if the function $S^{}_{xx}(t)$
depends on time as $A-B\ln|t|$, which has a Fourier transform
$B/(2f)$ at $f\neq 0$. The correlation function $S^{}_{xx}(t)$ can
be represented using the expansion near $t^{}_{f}\approx 1/(2\pi
f)$ in the approximate form $S^{}_{xx}(t)\approx
S^{}_{xx}(t_{f})+\frac{dS^{}_{xx}(t^{}_{f})}{d\ln
t^{}_{f}}\ln(t/t^{}_{f})$ (higher order expansion terms are
smaller by the factor $\ln^{-1}(AT^3/f) \approx 0.1$ for the small
frequency of interest $f \approx 0.1$Hz). The $1/f$ noise power
spectral density can then be expressed as
$S^{}_{xx}(f)=-(1/2f)dS^{}_{xx}(t^{}_{f})/d\ln t^{}_{f}$.

\begin{figure}[h!]
\centering
%\subfloat[]{
\includegraphics[scale=0.35]{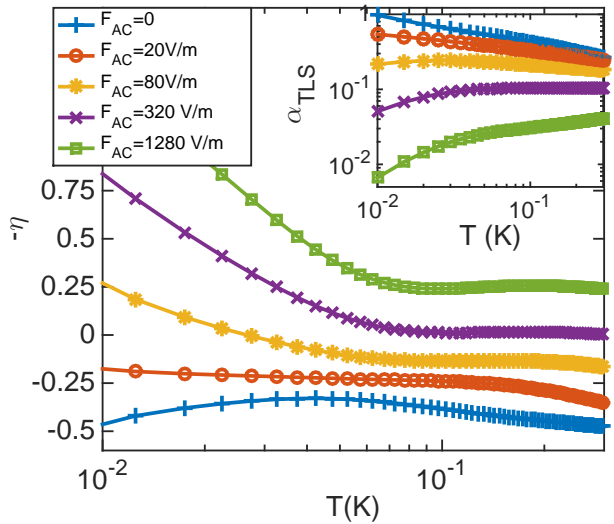}
%}%\hspace{10pt}
%\subfloat[]{\includegraphics[scale=0.3]{ExponentvsT.eps}}
{\caption{\small The exponent $\eta$ in the noise temperature
dependence $\alpha^{}_{TLS}\propto T^{-\eta}$ and the noise
parameter $\alpha^{}_{TLS}$ (inset) vs. temperature. TLS
parameters are chosen as\cite{abEcho} $p^{}_{0}=5$D,
$P_{0}U_{rms}=5\cdot 10^{-4}$, $A=10^8$s$^{-1}$K$^{-3}$,
$\omega=2\pi\cdot 5$GHz. \label{fig:NoisevsT1}}}
\end{figure}

For a quantitative comparison of the theory with the experiment it
is convenient to introduce a volume independent parameter in a
similar way to the experimentally determined Hooge's constant for
$1/f$ conductivity noise in semiconductors.\cite{Hooge} We define
this parameter as
\begin{widetext}
\begin{eqnarray}
\alpha^{}_{TLS}\equiv\frac{P^{}_{0}Vk^{}_{B}TfS^{}_{xx}(f)}{\tan^{2}\delta}=\frac{9\pi}{8\langle
p^{2}_{0}\rangle^2\hbar^2}
\int_{0}^{1}\frac{dx}{x(1-x^2)}\int_{0}^{1}dy\Bigg\langle\frac{p_{0}^4x^4y^4\chi
(k^{}_{B}T)^2} { \left[3W^{}_{0}(1/(2\pi f))+
\frac{T^{-1}_{2}(x)}{\sqrt{1-x^2}}
\sqrt{1+(xyp^{}_{0}F^{}_{AC})^2T^{}_{1}(x)T^{}_{2}(x)/\hbar^{2}}
\right]^2}\Bigg\rangle, \label{eq:noise4}
\end{eqnarray}
\end{widetext}
which is the ratio of the noise spectral density
multiplied by the number of thermal TLSs,
$N^{}_{T}=P^{}_{0}Vk^{}_{B}T$, to the squared average loss tangent
due to the TLSs,
$\tan\delta=\langle\epsilon''\rangle/\epsilon=\left(4\pi^2/3\epsilon\right)P^{}_{0}\langle
p^{2}_{0}\rangle\tanh\left(\hbar\omega/2k_{B}T\right)$. One should
notice the logarithmic weak frequency dependence of the noise
amplitude equivalent to a hardly distinguishable $f^{0.1}$
dependence for the typical system parameters (see  the caption to Fig.\
\ref{fig:NoisevsT1}). 

%minimum TLS tunneling amplitude $\Delta_{0min}$ following Ref. \cite{Faoro3} (lower limit $r_{min}=\Delta_{0min}/(\hbar\omega)$ in the integral in Eq. (\ref{eq:distr})) this will replace a $1/f$ dependence of the noise power with the dependence $\frac{{\rm atan}(f\tau_{max})}{f}$ ($\tau_{max} \approx (AT^3 r_{min}^2)^{-1}$ is the maximum TLS relaxation time). Therefore $1/f$ noise disappears in the regime $f\tau_{max} \gg 1$ studied in Ref. \cite{Faoro3}.

Consider the noise amplitude temperature dependence in the linear
regime $F^{}_{AC}\rightarrow 0$. For typical TLS parameters
$\chi\approx 5\cdot 10^{-4}$ and $A\approx
10^{8}$s$^{-1}$K$^{-3}$,\cite{abEcho} one has $W_{0}(T)\sim
10^{8}T$(s$^{-1}$), $T^{-1}_{2}\approx 10^{9}\,T^{2}$(s$^{-1}$)
and $T^{-1}_{1}\approx 10^{6}$s$^{-1}$. Consequently, the second
term in the denominator of the integrand in Eq.\ \eqref{eq:noise4}
can be approximately neglected at temperatures $0.01$K$<T<0.1$K.
The temperature dependence of the noise amplitude can thus be
approximated by $S^{}_{xx}(f)\propto T^{-1}\ln^{-2}\left(3.3\cdot
AT^3/2\pi f\right)$. The approximate exponent in the power law
temperature dependence $S^{}_{xx}(f)\propto T^{-1-\eta}$ observed
in the experiment can be estimated as $\eta=-1-d\ln
S^{}_{xx}(f)/d\ln T=-d\ln \alpha^{}_{TLS}/d\ln
T=6/\ln\left(3.3\cdot AT^3/2\pi f\right)$.\cite{abNoise} Setting
$f=0.1$Hz,\cite{NoiseRecent} we obtain $\eta\approx 0.42$, in
agreement with the experimental observation. The actual dependence
is slightly weaker because of the logarithmic integral in Eq.\
\eqref{eq:noise4}. Thus, the reported anomalous temperature
dependence of the $1/f$ noise amplitude in superconducting
resonators at low temperatures can be interpreted within the
framework of the standard tunneling model.

% Parameters to try
% Experiment

The temperature dependence of the noise parameter
$\alpha^{}_{TLS}$ [Eq.\ \eqref{eq:noise4}] and the exponent $\eta$
extracted from it are shown in Fig.\ \ref{fig:NoisevsT1} for
various external field amplitudes. It is evident that in the
linear regime, $F_{AC}\rightarrow 0$, the temperature dependence
observed in the experiment in the regime $k^{}_{B}T\ll\hbar\omega$
is qualitatively reproduced. The accurate comparison of the theory
with the experiment is the subject of a separate work that will
help to determine the system parameters from the optimum data fit.

It should be emphasized that our results do not rule
out the possibility of some energy dependence of the DOS. For the
specific case of $1/f$ noise it shows that a significant
contribution to the exponent $\eta$ may arise due to the broad
spectrum of relaxation times in amorphous systems, even if the DOS
is energy independent. However, other deviations from the STM,
such as the non-integer exponents of the specific heat and thermal
conductivity may originate from an energy dependence of the DOS at
low energies. Further investigation is thus necessary in order to
shed light on the energy dependence of the DOS. This can be
achieved by additional measurements, such as temperature and
frequency dependent dielectric losses or internal friction in the
plateau regime, which should be both sensitive to the energy
dependence of the DOS.\cite{Hunklinger}

%The presence of some dependence of TLS density on energy of course
%cannot be completely excluded as it is seen in the deviations of
%the temperature dependence of TLS thermal conductivity of the
%$T^{2}$ law \cite{YuLeggett} and can be the consequence of the
%specific nature of observed two level systems \cite{Moshe}.
%Yet, this can be verified by more straightforward measurements
%including temperature and frequency dependent dielectric losses or
%internal friction in the plateau regime which should be both
%sensitive to that energy dependence.\cite{Hunklinger}

The strong reduction of the noise with increasing field amplitude,
as well as the weakening of its temperature dependence
 are clearly seen in Fig.\ \ref{fig:NoisevsT1}.
These findings are in agreement with the
experimental results.\cite{NoiseRecent} Yet, in the large field
limit, $F^{}_{AC}>\left(\hbar
W^{}_{0}(t)/p^{}_{0}\right)\sqrt{T^{}_{2}/T^{}_{1}}\approx
1000$V$/$m, Eq.\ \eqref{eq:noise4} predicts $S^{}_{xx}\propto
TF^{-2}_{AC}$, while the observed field dependence is
weaker.\cite{NoiseRecent} This might be due to nonequilibrium
heating\cite{McCammon} and insufficiently large  AC
field used experimentally. A detailed analysis requires
understanding of the mechanisms of heat exchange between the
superconducting circuit and the environment, which is beyond the
scope of this letter.

% Remake a graph with inset containing exponent vs temperature
% Discuss temperature and field dependencies, heating effect
% Discuss high temeprature: fail of both models and TLS relaxation, anomalous T2 behavior and quasi particles
% Conclusions, request graph from Moshe
% Check literature, rearrange and make it according to the rules

At high temperatures, $k^{}_{B}T>\hbar\omega$, Eq.\
\eqref{eq:noise4} as well as the earlier
work\cite{NoiseRecent,Faoro3} predict the strong reduction of the
noise with increasing temperature due to the factor
$\tanh^2\left(E^{}_{i}/2k_{B}T\right)\propto T^{-2}$. Such
temperature dependence does not fit the experimental
observation,\cite{NoiseRecent} which reports a weaker temperature
dependence. We expect that different mechanism may be responsible
for the $1/f$ noise in this regime. According to our preliminary
analysis this cannot be the relaxational interaction of TLS with
the AC field\cite{Hunklinger} since it does not lead to $1/f$
noise. Indeed, the low frequency noise should be due to slow TLSs
with the large relaxation times $T^{}_{1}\approx 1/2\pi f$. They
can indeed contribute to the fluctuations of a zero frequency
dielectric constant\cite{Clare0} while their response to an
 AC field vanishes as $f/\omega$.\cite{Hunklinger}

\begin{figure}[h!]
\centering
%\subfloat[]{
\includegraphics[scale=0.3]{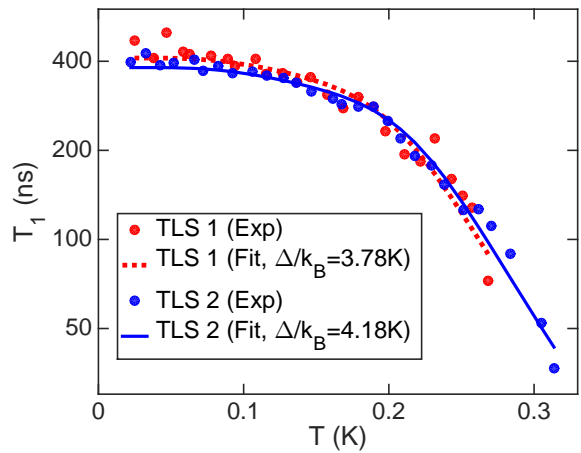}
%}%\hspace{10pt}
%\subfloat[]{\includegraphics[scale=0.3]{ExponentvsT.eps}}
{\caption{\small Experimental data\cite{UstinovRecent} for
temperature dependent relaxation times of two different TLSs
interpreted assuming the presence of quasi-particles.
\label{fig:QuasiPart}}}
\end{figure}

Alternatively, the noise in dielectric tunneling barrier
incorporated in superconducting devices may be due to
quasi-particle  excitations. In particular, a recent study has
revealed an anomalous temperature dependence of the relaxation and
decoherence rates of single TLSs in the amorphous tunnel barrier
of a superconducting phase qubit, showing a deviation from the
single-phonon relaxation rate [Eq.\ \eqref{eq:TLSRelRate}] at
temperatures above 0.1K.\cite{UstinovRecent} As shown in Fig.\
\ref{fig:QuasiPart}, the relaxation rate of these TLSs can be
fitted by combining the contributions to the relaxation due to
TLS-phonon interaction and the thermally activated quasi-particle contribution\cite{BF} as
$T^{-1}_{1}(T)=A\coth\left(E/2k^{}_{B}T\right)+Be^{-\Delta/2k^{}_{B}T}$.
For both TLSs studied in Ref.\ \onlinecite{UstinovRecent}, the
fitting parameter $\Delta$ is close to the Al superconducting gap
$3.95k^{}_{B}$K. These quasi-particles may also be responsible for
$1/f$ noise at temperatures exceeding $0.1$K.

In conclusion, we calculated the $1/f$ noise in
microwave dielectric constant produced by TLSs in amorphous
solids. The STM involving the long range TLS-TLS interactions has
been used. The results are consistent with the recent experimental
data at low temperature, whereas at higher temperature other
mechanism may be responsible for the noise, possibly one
associated with broken Cooper pairs.

We acknowledge Kevin Osborn and Georg Weiss for stimulating
discussions and Louisiana EPSCORE LA Sigma and LINK Programs for
the support.

\end{document}